\documentclass[aps,prl,preprint,groupedaddress]{revtex4}

\usepackage{graphicx}%
\usepackage{bm}

\begin{document}

% Use the \preprint command to place your local institutional report
% number in the upper righthand corner of the title page in preprint mode.
% Multiple \preprint commands are allowed.
% Use the 'preprintnumbers' class option to override journal defaults
% to display numbers if necessary
%\preprint{}

\title{
  Spin-flop ordering from frustrated ferro-/antiferro-magnetic interactions:
  \\ a combined theoretical and experimental study of a Mn/Fe(100)
  monolayer
}

% repeat the \author .. \affiliation  etc. as needed
% \email, \thanks, \homepage, \altaffiliation all apply to the current
% author. Explanatory text should go in the []'s, actual e-mail
% address or url should go in the {}'s for \email and \homepage.
% Please use the appropriate macro foreach each type of information

% \affiliation command applies to all authors since the last
% \affiliation command. The \affiliation command should follow the
% other information
% \affiliation can be followed by \email, \homepage, \thanks as well.

\author{C. Grazioli}
\affiliation{Istituto di Struttura della Materia, Consiglio Nazionale
  delle Ricerche, Area Science Park, I-34012 Trieste, Italy} 

\author{Dario Alf\`{e}} 
\affiliation{ 
  Department of Earth Sciences and Department of Physics and
  Astronomy, University College London, Gower Street, London, WC1E
  6BT, UK } 
\affiliation{
  INFM DEMOCRITOS National Simulation Center, Trieste, Italy} 

\author{S.R. Krishnakumar}
%\email[]{Your e-mail address}
%\homepage[]{Your web page}
%\thanks{}
%\altaffiliation{}
\altaffiliation{
Present address: Surface Physics Division, Saha Institute of Nuclear Physics, Kolkata - 700 064,
India.
}
\affiliation{
  International Centre for Theoretical
Physics, Strada Costiera 11, 34100 Trieste, Italy} 

\author{Subhra Sen Gupta}
%\email[]{Your e-mail address}
%\homepage[]{Your web page}
%\thanks{}
%\altaffiliation{}
\altaffiliation{Also at Department of Physics, Indian Institute of Science, Bangalore 560012, India.}
\affiliation{Solid State and Structural Chemistry Unit, Indian
  Institute of Science, Bangalore 560 012, India} 

\author{M. Veronese}
%\email[]{Your e-mail address}
%\homepage[]{Your web page}
%\thanks{}
%\altaffiliation{}
\affiliation{Istituto di Struttura della Materia, Consiglio Nazionale
  delle Ricerche, Area Science Park, I-34012 Trieste, Italy} 

\author{S. Turchini}
%\email[]{Your e-mail address}
%\homepage[]{Your web page}
%\thanks{}
%\altaffiliation{}
\affiliation{Istituto di Struttura della Materia, Consiglio Nazionale
  delle Ricerche, Area Science Park, I-34012 Trieste, Italy} 

\author{Nicola Bonini}
\affiliation{SISSA -- Scuola Internazionale Superiore di Studi
  Avanzati, Via Beirut 2-4,  34014 Trieste, Italy}
\affiliation{
  INFM DEMOCRITOS National Simulation Center, Trieste, Italy} 

\author{Andrea Dal Corso}
\affiliation{SISSA -- Scuola Internazionale Superiore di Studi
  Avanzati, Via Beirut 2-4,  34014 Trieste, Italy}
\affiliation{
  INFM DEMOCRITOS National Simulation Center, Trieste, Italy} 

\author{D.D. Sarma}
%\email[]{Your e-mail address}
%\homepage[]{Your web page}
%\thanks{}
%\altaffiliation{}
\affiliation{Solid State and Structural Chemistry Unit, Indian
  Institute of Science, Bangalore 560 012, India} 

\author{Stefano Baroni}
%\email[]{Your e-mail address}
%\homepage[]{Your web page}
%\thanks{}
%\altaffiliation{}
\affiliation{SISSA -- Scuola Internazionale Superiore di Studi
  Avanzati, Via Beirut 2-4,  34014 Trieste, Italy}
\affiliation{
  INFM DEMOCRITOS National Simulation Center, Trieste, Italy} 

\author{C. Carbone}
%\email[]{Your e-mail address}
%\homepage[]{Your web page}
%\thanks{}
%\altaffiliation{}
\affiliation{Istituto di Struttura della Materia, Consiglio Nazionale
  delle Ricerche, Area Science Park, I-34012 Trieste, Italy}

%Collaboration name if desired (requires use of superscriptaddress
%option in \documentclass). \noaffiliation is required (may also be
%used with the \author command).
%\collaboration can be followed by \email, \homepage, \thanks as well.
%\collaboration{}
%\noaffiliation

\date{\today}

\begin{abstract}
  The occurrence of a non-collinear magnetic structure at a Mn
  monolayer grown epitaxially on Fe(100) is predicted theoretically,
  using spinor density-functional theory, and observed experimentally,
  using x-ray magnetic circular dichroism (XMCD) and linear dichroism
  (XMLD) spectroscopies. The combined use of XMCD and XMLD at the Mn-
  absorption edge allows us to assess the existence
  of ferro-magnetic and antiferro-magnetic order at the interface, and
  also to determine the moment orientations with element specificity.
  The experimental results thus obtained are in excellent agreement
  with the magnetic structure determined theoretically.
\end{abstract}

% insert suggested PACS numbers in braces on next line
% insert suggested keywords - APS authors don't need to do this
%\keywords{}

%\maketitle must follow title, authors, abstract, \pacs, and \keywords
\maketitle

% body of paper here - Use proper section commands
% References should be done using the \cite, \ref, and \label commands

The magnetic structure of ultra-thin anti-ferromagnetic (AFM)
over-layers on ferromagnetic (FM) substrates determines the properties
of ferro-antiferro magnetic multi-layers which are key constituents of
devices such as exchange-bias or tunnelling magneto-resistance
recording systems. Complex, non-collinear, magnetic structures are
expected at these interfaces, for spin canting minimizes the exchange
energy between a ferromagnet and an anti-ferromagnet that exposes a
plane with anti-parallel spins \cite{Hinchey}. The resulting {\em
  spin-flop} alignment of the moments in the anti-ferromagnet,
perpendicular to the magnetization in the ferromagnet, is the 
microscopic basis of the large coercive field in exchange-bias devices 
\cite{Koon,Schulthess}. In spite of the great interest in these
systems, the understanding gained so far on the basis of
semi-empirical models has not been validated by accurate
first-principles calculations, nor by a direct experimental
observation of the non-collinear magnetic order at the
interface. In fact, on one hand, accurate, fully unconstrained, methods
based on density-functional theory (DFT) for studying non-collinear
magnetic structures have become available only recently 
\cite{Oda,Ralph,vasp-NC}; on the other hand, the simultaneous access
to both antiferromagnetic and ferromagnetic ordering, not possible by usual experimental
methods, has become possible by magnetic circular and linear
dichroism methods, using advanced synchrotron sources.

Thin films of Mn on Fe(100) represent an interesting case where 
unusual magnetic structures can occur. For Mn coverages larger than 2
mono-layers (ML), an anti-ferromagnetic coupling between adjacent Mn
layers was found by spin-polarized electron energy loss spectroscopy 
\cite{Walker} and by spin-polarized scanning tunneling spectroscopy
and microscopy \cite{Schlickum,Yamada}. In the monolayer and 
submonolayer regimes the magnetic structure is expected to be 
dominated by the frustration arising between competing ferro- and
antiferro-magnetic interactions.  DFT calculations indicate that Mn-Mn
interactions in the over-layer evolve from FM in the diluted
(low-coverage) limit \cite{Nonas} to AFM at a coverage of 1ML 
\cite{Wu}. This behavior was observed experimentally using 
spin-resolved core level photo-emission \cite{Roth} and confirmed by 
x-ray magnetic circular dichroism \cite{Rader}. The possible 
occurrence of a non-collinear spin order at the interface was 
suggested on the basis of simplified DFT calculations \cite{Spisak}.

The aim of this work is to determine the magnetic structure of Mn 
deposited on Fe(100) in the monolayer regime, using state-of-the-art 
theoretical and experimental techniques. Mn/Fe(100) has been simulated 
using fully unconstrained Spinor Density Functional Theory (SDFT) 
which allows for a proper account of non-collinear magnetic structures 
\cite{Oda,Ralph,vasp-NC}.  This same system has then been investigated 
experimentally by means of a combination of x-ray magnetic circular 
dichroism (XMCD) and x-ray magnetic linear dichroism (XMLD) 
spectroscopies, which allows for a direct assessment of ferromagnetic 
and antiferromagnetic ordering, and also for a determination of the moment 
orientations with chemical sensitivity. 

Our SDFT calculations have been performed within the local density 
approximation \cite{dft-details} and neglecting spin-orbit 
interactions. The latter approximation makes the calculated magnetic 
structures degenerate with respect to an arbitrary overall rotation of 
the magnetization field. At low pressure and temperature iron has a 
ferromagnetic body-centered-cubic structure, while manganese is 
orthorhombic with a complex antiferromagnetic order.  It turns out 
that the equilibrium structure of a Mn monolayer is also 
anti-ferromagnetic. By constraining all the magnetic moments to be
collinear (which amounts to performing a conventional local
spin-density calculation) a Mn overlayer on Fe can assume one of three
distinct configurations: in the {\em ferromagnetic} (FM) structure all
the Mn magnetic moments are parallel to each other and to the
magnetization of the ferromagnetic iron substrate; in the second,
{\em anti-ferromagnetic} (AFM), structure the Mn magnetic moments are
antiparallel with respect to the substrate magnetization; the third,
{\em ferrimagnetic} (FI), structure is finally characterized by an
anti-ferromagnetic arrangement of atomic moments in the overlayer,
which result alternatively parallel or anti-parallel to
the substrate magnetization. It is interesting to notice that the
inequivalence of spin-up and spin-down atoms in the overlayer may
determine some buckling in the surface.

\begin{figure} [t!]
\includegraphics[width=10.0cm]{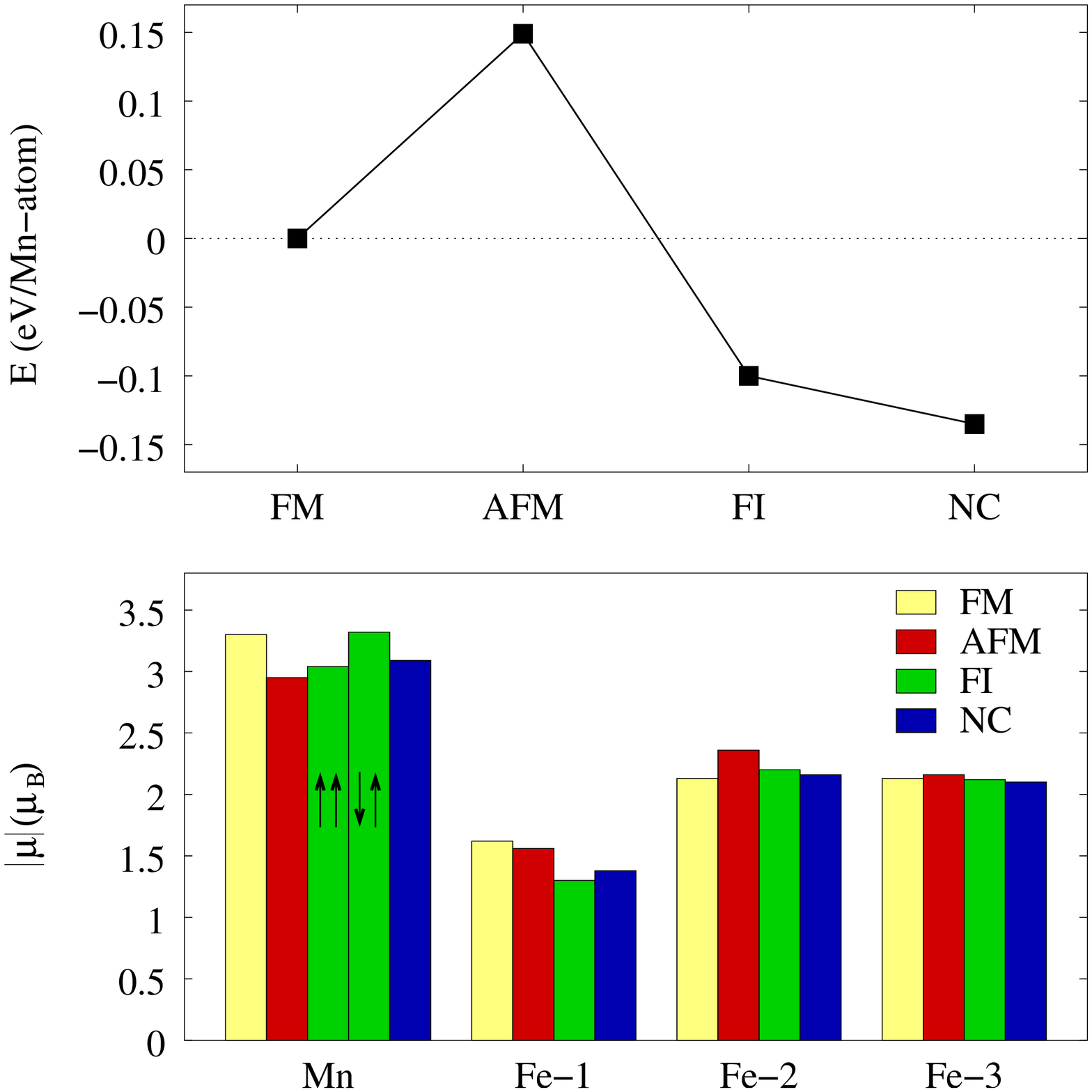}
\caption{\label{fig1} Top panel: total energies of the FM, AFM, FI and NC 
structures relative to the FM structure. Bottom panel: magnitude of the 
atomic magnetic moments in the various structures. For the FI structure
$\uparrow \uparrow$ and $\uparrow \downarrow$ indicate the Mn atom with 
magnetic moment parallel and antiparallel to the underlying Fe atoms 
respectively. }
\end{figure}

In Fig. \ref{fig1} (top) we report the energies of the various magnetic
structures relative to the energy of the FM structure. As a
consequence of the stability of the anti-ferromagnetic order of an
isolated Mn mono-layer, and in agreement with previous studies 
\cite{Wu}, we find that among these three structures the FI one is the
most stable.  The largest stability of this structure implies that
half of the Fe-Mn magnetic bonds across the interface is frustrated,
whether or not these bonds are preferentially ferro- or
antiferro-magnetic.  The FM structure is actually slightly more stable
than the AFM one, indicating that the exchange interaction between Mn
and Fe atoms is preferentially ferromagnetic. In Fig. \ref{fig1} (bottom) we
report the magnitude of the calculated atomic moments, defined as the
integral of the magnetization inside a sphere centered on the atoms
and with a radius equal to the nearest neighbor distance. We note that
in the FI structure the Mn surface layer has nearly zero total
magnetization, in accordance with the experimental findings of
Ref. \cite{Rader}.

\begin{figure} [t!]
\includegraphics[width=10.0cm]{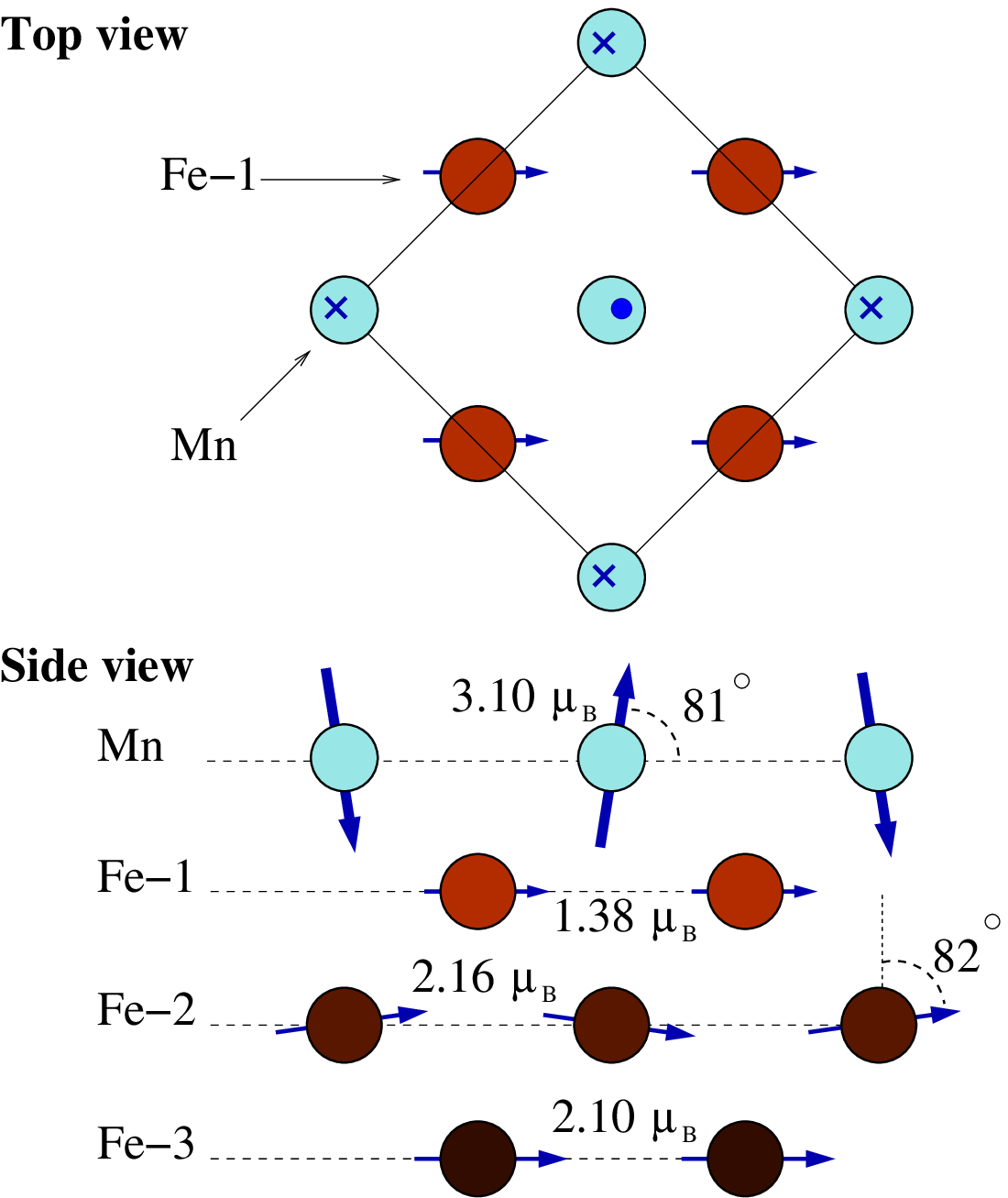}
\caption{\label{fig2}Schematic view of the Mn/Fe magnetic interface.}
\end{figure}

When the constraint of spin collinearity is released, the frustration
of the magnetic bonds across the interface drives a rotation of the Mn
moments resulting in a chess-board arrangement where these moments
form angles of $\approx\pm 80^\circ$ with respect to those of the
underlying Fe atoms.\cite{NC-stability} This non-collinear (NC)
structure is depicted in Fig. \ref{fig2}. We note that, with such arrangement,
the magnetic interactions between Mn and Fe are the same for all the Mn
atoms, which are therefore all structurally equivalent. The almost
perpendicular orientation of the Mn moments corresponds to a
quasi-anti-ferromagnetic order in the Mn plane, still avoiding the
magnetic frustration experienced in the collinear situations. We find
that this NC structure is more stable by about 35 meV/atom than the FI
collinear structure (see the top panel of Fig. \ref{fig1}), and should therefore be
clearly observable also at room temperature. The departure from a
$90^\circ$ orientation is small, but possibly significant. In fact, as
reported above, among the collinear structures the FM one is slightly
more favorable than the AFM one, and this may be the cause of the
small ferromagnetic bias in the NC structure.

Using XMCD and XMLD at the Mn-$L_{2,3}$ edges we examined the 
magnetic structure of sub-monolayer and monolayer (0.1-1.0 ML) 
Mn films deposited on Fe(100). The samples were grown \emph{in situ} 
in the $10^{\mathrm{{-10}}}$ mbar range using a quartz-crystal microbalance
to control the thickness. 
As a first step, an Fe(100) surface was epitaxially
grown on a clean and ordered Ag(100) single crystal surface \cite{jonker} 
and magnetized in remanence along the Fe[001]. 
The thickness of the  Fe film was above 50 ML in order to suppress 
Ag surface segregation. 
A wedge of Mn in
the range of 0-1.1 ML was prepared. The experiments were performed 
at the 4.2 beamline "Circular Polarization" at the ELETTRA
storage ring, using approximately 95$\%$ linearly or 70$\%$ circularly
polarized light and were 
collected in the total electron yield (TEY) mode.

\begin{figure} [t]
\includegraphics[width=10.0cm]{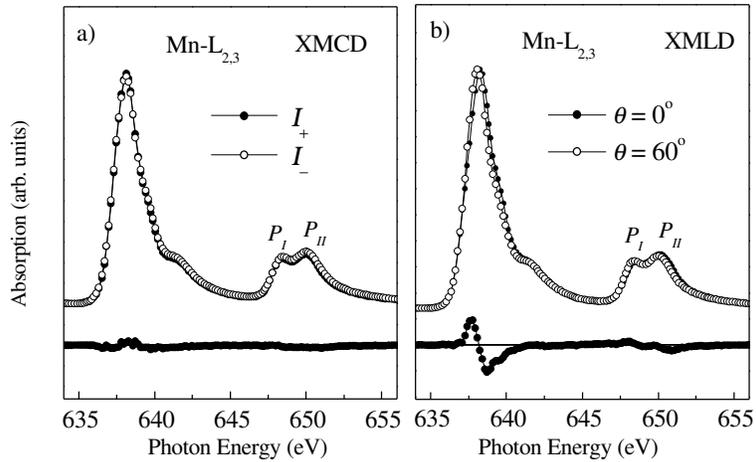} 
\caption{\label{fig3} a) XMCD data at the $L_{2,3}$ edges for 1 ML of
  Mn. b) Mn-$L_{2,3}$ XAS for 1 ML of Mn as a function of the angle
  $\theta$ between $\bm{E}$  
  and the surface normal. The difference spectra are plotted at the
  bottom.} 
\end{figure}

The magnetization of Mn was investigated by circular dichroism. The
XMCD data measurements performed at different Mn thicknesses for the
submonolayer regime (not shown here) are in accordance with Rader \emph{et al.} \cite{Rader} and
Dresselhaus \emph{et al.} \cite{Dresselhaus}. For less than 1 ML we observe an XMCD effect of the
Mn adlayer opposite to that of Fe, which is a proof of long-range
ferromagnetic order of the adlayer aligned antiparallel to the
magnetization in the Fe substrate. The net magnetization of Mn
decreases with increasing thickness and approaches zero for 1 ML, thus
pointing out a rather smooth transition from ferro- to
antiferromagnetic arrangement. The XMCD data measured on 1ML are reported in
Fig. \ref{fig3}a.

\begin{figure} [t]
\includegraphics[width=10.0cm]{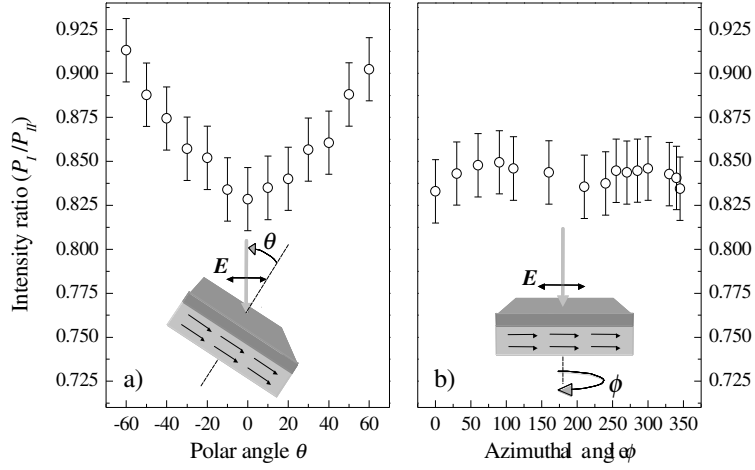} 
\caption{\label{fig4} XMLD effect for the polar (a) and for the
  azimuthal (b) rotations. The intensity ratio of the Mn-$L_{2}$ double
  peak structures is used to measure the intensity of the effect.} 
\end{figure}

The XMLD was obtained keeping the direction of the electric vector
$\bm{E}$ of the incident linearly polarized light fixed in space and
rotating the sample as illustrated schematically in Fig. \ref{fig4}. 
Since the XMLD effect is maximized when
the magnetization is switched between parallel and perpendicular to
the photon polarization, the sample was rotated in the
polar geometry (Fig. \ref{fig4}a). Finally the sample was rotated in the azimuthal geometry
(Fig. \ref{fig4}b).
The absorption edges recorded for in-plane and for
out-of-plane polarization (polar rotation) are shown in
Fig. \ref{fig3}b. The features of these spectra can be identified 
with Mn-$3d^5$ multiplet structures in a high-spin state, suggesting 
a local magnetic moment larger than $3.5\mu_{B}$. 
We note that DFT calculations may underestimate 
the magnetic moment in the case of Mn \cite{raderobrien}. 
%From the comparison we can notice that the structures of the spectrum change
%their relative intensities, that can be noticed particularly well in
%the structure at 650.3 for $L_{2}$; secondly, each spectral
%feature is shifted to higher energies when we rotate the photon
%polarization out of plane.

The intensity ratio between the $L_{2}$ well defined double peak structures
(labeled as $P_{I}$ and $P_{II}$) turns out to be well suited to
detect the spectral changes. 
The intensity ratios for the two rotation geometries, reported in
Fig. \ref{fig4}, show that the polar measurement
has a regular dependence as a function of the angle, while the azimuthal
one is much less pronounced. This indicates that the magnetic moments have preferential alignment with respect to the surface plane. The analysis of the $P_{I}/P_{II}$ intensity ratios, as discussed below, supports the idea that the orientation is perpendicular to the surface \cite{inplane}.

We have simulated the spectral lineshapes by exact diagonalization of an
atomic many-body Hamiltonian, based on a fully coherent spectral
function given by: 
%\begin{equation}
%I_{q}(\omega)=
%\sum_{n}\left|\sum_{v,c}\langle\Psi_{f}^{n}(N)|a_{v}^{\dagger}a_{c}|
%\Psi_{G}(N)\rangle \langle v|rC^{(1)}_{q}|c\rangle\right|^{2}
%\delta(\omega-(E_{f}^{n}(N)-E_{g}(N)))
%\end{equation}
$$ \begin{array}{ll}
 I_{q}(\omega)= & \sum_{n}\left|\sum_{v,c}\langle\Psi_{f}^{n}(N)|a_{v}^{\dagger}a_{c}|\Psi_{G}(N)\rangle \langle v|rC^{(1)}_{q}|c\rangle\right|^{2} \\
& \delta(\hbar\omega-(E_{f}^{n}(N)-E_{g}(N))). \\
 \end{array}$$
The irreducible components of the dipole operator are
defined as: $T^{(1)}_{q} = rC^{(1)}_{q}$, 
where $q = 0,+1,-1$ correspond to z-linearly polarized, 
right circularly polarized and left circularly polarized light, 
respectively, and $C^{(1)}_{q}$ are normalized spherical harmonics \cite{simuldetails,dd_subhra}.

\begin{figure} [t]
\includegraphics[width=10.0cm]{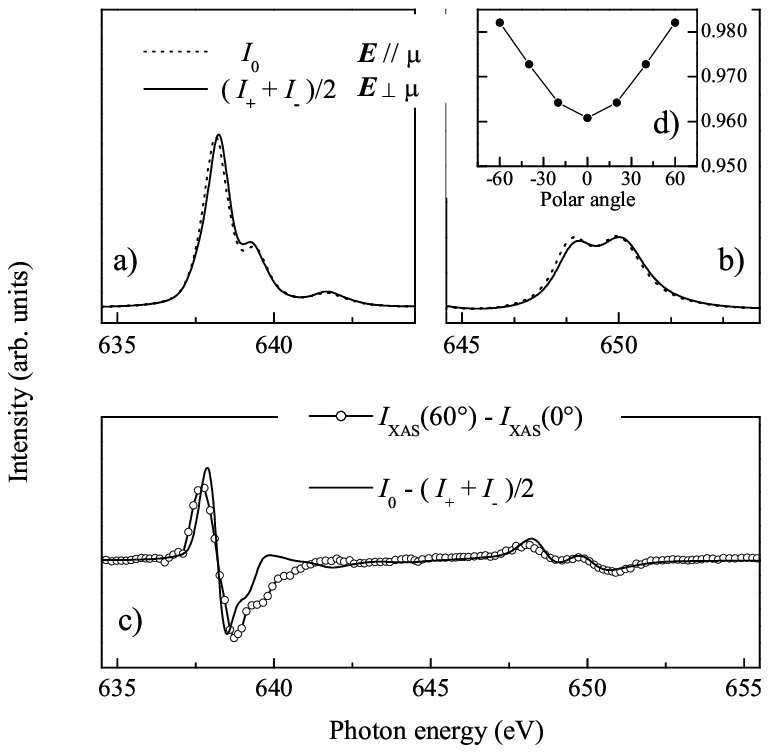} 
\caption{\label{fig5} a) and b) Simulation of the absorption spectra 
for the two polarization directions after suitable broadening.  
c) comparison between the simulated and the experimental XMLD. 
d) simulated $P_{I}/P_{II}$ intensity ratio.
}
\end{figure}

In Fig.~\ref{fig5}a,b we show the results of the spectral simulations. 
By using different combinations of the irreducible components of the 
dipole operator $T^{(1)}_{q}$ we are able to simulate 
the Mn-$L_{2,3}$ spectra as a function of the orientation between the 
magnetic moment on Mn and $\bm{E}$: 
$I_{0}$ for $\bm{E}\parallel\bm{\mu}$ and 
$\frac{1}{2}(I_{+} + I_{-})$ for $\bm{E}\bot\bm{\mu}$. 
In terms of the lineshape, the agreement with the experimental spectra of 
Fig. \ref{fig3}b is very good. Remarkably, also the calculated spectra 
display the lineshape modifications (changes in the relative peak intensities and 
relative peak energy positions) that have been detected experimentally. 
As shown in Fig. \ref{fig5}c, the XMLD spectrum, which carries all information about 
the lineshape changes, is well reproduced by the simulated XMLD 
lineshape calculated as $I_{0} - \frac{1}{2}(I_{+} + I_{-})$ 
The calculated $P_{I}/P_{II}$ intensity ratio, assuming perpendicular orientation of the Mn moments with respect to the surface plane, also simulates the angular dependence of the experimental results. 
We can conclude that the magnetic moments of Mn are aligned
out of the surface plane, perpendicular to the Fe magnetization.

It is known from the experimental works in Ref.~\cite{Schlickum} and \cite{Yamada} that in thicker Mn(100) films the layers are coupled antiferromagnetically to each other. The growth of Mn over a step edge of the substrate gives rise to a topological frustration among adjacent layers with opposite magnetization direction \cite{Schlickum}. The magnetic frustration is, in such a case, relaxed through the formation of a narrow and magnetically non-collinear structure, similar to a $180\,^{\circ}$ domain wall, with lateral extension of the order of a few nanometers. In the present case, instead, the non-collinear magnetism of the system intrinsically derives from the competing exchange interactions between the monolayer and the substrate moments, and therefore uniformly extends over a macroscopic area. This mechanism may also lead to the development of similar spin-flop structures in other monolayer systems, where the exchange interaction within the antiferromagnetic monolayers prevails, but does not overwhelm, the coupling with the ferromagnetic substrate.

D.A. would like to thank the Royal Society and the Leverhulme Trust for support. S.R.K. would like to thank ICTP for TRIL Fellowship.

% Create the reference section using BibTeX:

\begin{thebibliography}{}

\bibitem{Hinchey} 
  L. L. Hinchey and D. L. Mills, Phys. Rev. B \textbf{34}, 1689 (1986).  

\bibitem{Koon} 
  N. C. Koon. Phys. Rev. Lett. \textbf{78}, 4865 (1997).

\bibitem{Schulthess} 
  T. C. Schulthess and W. H. Butler, Phys. Rev. Lett. \textbf{81}, 4516 (1998). 

\bibitem{Oda}
  T. Oda, A. Pasquarello, and R. Car, Phys. Rev. Lett. {\bf 80}, 3622 (1998).

\bibitem{Ralph}
  R. Gebauer \emph{et al.}, Phys. Rev. B {\bf 61}, 6145 (2000).

\bibitem{vasp-NC}
  D. Hobbs, G. Kresse, and J. Hafner, Phys. Rev. B {\bf 62}, 11556 (2000).

\bibitem{Walker} 
  T. G. Walker and H. Hopster, Phys. Rev. B \textbf{48}, R3563 (1993).

\bibitem{Schlickum} 
  U. Schlickum, N. Janke-Gilman, W. Wulfhekel, and J. Kirschner, Phys. Rev. Lett. \textbf{92}, 107203 (2004). 
  
\bibitem{Yamada} 
  T. K. Yamada, M. M. J. Bischoff, G. M. M. Heijnen, T. Mizoguchi, and H. van Kempen, Phys. Rev. Lett. \textbf{90}, 056803 (2003). 

\bibitem{Nonas} 
  B. Nonas, K. Wildberger, R. Zeller, and P. H. Dederichs, Phys. Rev. Lett. \textbf{80}, 4574 (1998).
  
\bibitem{Wu} 
  R. Wu and A. J. Freeman, Phys. Rev. B \textbf{51}, 17131 (1995).

\bibitem{Roth} 
  Ch. Roth, T. Kleemann,F. U. Hillebrecht, and E. Kisker, Phys. Rev. B \textbf{52}, R15691 (1995). 

\bibitem{Rader} 
  O. Rader, W. Gudat,D. Schmitz,C. Carbone, and W. Eberhardt, Phys. Rev. B \textbf{56}, 5053 (1997). 
  
\bibitem{Spisak} 
  D. Spi\v s\`ak and J. Hafner, Phys. Rev. B \textbf{55}, 8304 (1997).

\bibitem{dft-details} 
  Our calculations have been performed using UltraSoft Vanderbilt 
  pseudopotentials~\cite{vanderbilt90} and plane-waves with a cut-off 
  of 408 eV. Surfaces have been modeled using symmetric slab geometry, with 7
  atomic layers of Fe and Mn adsorbates on both sides of the slab, and
  a vacuum region corresponding to 6 atomic layers. The primitive cell
  used in our calculations contained two surface atoms, with 
  $c(2\times 2)$ geometry. Integration inside the Brillouin zone has been
  performed by summation over 18 uniformly spaced surface {\bf
  k}-points, and a smearing functions of width 0.34~eV.
  Calculations performed using 72 k-points and/or a smearing function 
  of 0.068 eV affected the magnitude of the calculated magnetic moments 
  by $\sim 0.02~\mu_B$ and their directions by a fraction of a degree

\bibitem{vanderbilt90} D. Vanderbilt, Phys. Rev. B {\bf 41}, 7892 (1990).

\bibitem{NC-stability} 
  We have performed several calculations starting from different
  initial magnetic configurations. All these
  calculations finally converged to either one of the three collinear
  magnetic structures described in the text, or to a non-collinear
  structure in which the Mn adlayer has an almost anti-ferromagnetic
  arrangement, as described in the text.
  
\bibitem{Dresselhaus} 
  J. Dresselhaus \emph{et al.}, Phys. Rev. B \textbf{56}, 5461 (1997).
  
\bibitem{jonker} 
  B. T. Jonker, K. H. Walker, E. Kisker, G. A. Prinz, and C. Carbone, Phys. Rev. Lett. 57, 142 (1986). 

\bibitem{raderobrien}
  O. Rader \emph{et al.}, Phys. Rev. B \textbf{55}, 5404 (1997); W.~L. O'Brien and B.~P. Tonner, Phys. Rev. B \textbf{51}, 617 (1995). 

\bibitem{inplane}
  The data suggest that there might be a little in-plane component 
  but the effect is too small compared to the error bar. 

\bibitem{simuldetails}
  The atomic many-body Hamiltonian involves the on-site energies of the
  Mn $2p$ and $3d$ levels, full multiplet Coulomb interaction within the
  Mn $3d$ manifold and between the Mn $2p$ and $3d$ manifolds and
  spin-orbit interaction in the Mn $3d$ and $2p$ levels \cite{dd_subhra}.
  Additionally we have used a Zeeman field (coupling to both the spin
  and orbital moments) of 7 Tesla and an exchange (spin-only) field of
  10 meV or 173 Tesla to simulate the exchange coupling between the
  magnetic ions in the system.
  The average Mn~$3d$-$3d$ multiplet interaction ($U_{dd}$) was chosen
  to be 4.0 eV.  The Slater-Condon multiplet parameters used in the
  calculation were taken from Ref. \onlinecite{vdL} and were
  scaled to 75\% of their atomic values to take into account screening
  effects of the metallic host. As is customary, the Mn~$2p$-$3d$
  average multiplet interaction ($U_{pd}$) was chosen to be $1.1U_{dd}=$ 4.4 eV. 

\bibitem{dd_subhra}
  P. Mahadevan and D.~D. Sarma, Phys. Rev. B {\bf 61}, 7402 (2000);
  Subhra Sen Gupta, P. Mahadevan and D.~D. Sarma (Unpublished
  results). 

\bibitem{vdL} 
  G. van der Laan \emph{et al.}, J. Phys.: Cond. Matter {\bf 4}, 4189 (1992). 

\end{thebibliography}
\end{document}